\documentclass[twocolumn]{aastex631}
\usepackage{array}    
\usepackage{booktabs} 

\begin{document}


\title{Mass-Dependent Radial Distribution of Single and Binary Stars in the Pleiades and their Dynamical Implications}

\author[0000-0001-5968-1144]{Rongrong Liu}
\affiliation{Shanghai Astronomical Observatory, Chinese Academy of Sciences, 80 Nandan Road, Shanghai 200030, People's Republic of China; zyshao@shao.ac.cn}
\affiliation{School of Astronomy and Space Sciences, University of Chinese Academy of Sciences, No. 19A Yuquan Road, Beijing 100049, People's Republic of China}

\author[0000-0001-8611-2465]{Zhengyi Shao}
\affiliation{Shanghai Astronomical Observatory, Chinese Academy of Sciences, 80 Nandan Road, Shanghai 200030, People's Republic of China; zyshao@shao.ac.cn}
\affiliation{Key Lab for Astrophysics, Shanghai 200234, People's Republic of China}

\author[0000-0002-0880-3380]{Lu Li}
\affiliation{Shanghai Astronomical Observatory, Chinese Academy of Sciences, 80 Nandan Road, Shanghai 200030, People's Republic of China; zyshao@shao.ac.cn}



\begin{abstract}

The Pleiades is a young open cluster that has not yet dynamically relaxed, making it an ideal target to observe various internal dynamical effects. By employing a well-defined sample of main-sequence (MS) cluster members, including both MS single stars and unresolved MS+MS binaries, we revisited their individual masses and mass functions and quantified the mass dependence of their radial distributions. We found that the mass function of binaries is more top-heavy than that of single stars. Significant mass segregation is observed for both single and binary populations respectively, with more massive objects concentrated towards the cluster center. Notably, within given mass ranges, binaries are distributed more scattered than single stars, providing direct evidence for more efficient dynamical disruption of binaries in the inner region. The radial distribution of the binary fraction, expressed as the $f_{\rm b}-R$ relation can be characterized by a bimodal shape, with higher $f_{\rm b}$ values in both innermost and outermost regions of the cluster. The lower-mass subsample exhibits a monotonic increase in $f_{\rm b}$ with radius, reflecting the impact of binary disruption. Conversely, for the higher-mass subsample, $f_{\rm b}$ decreases with radius. It can be explained that these massive cluster members, which possess higher binary probabilities, have already undergone significant mass segregation. All these observational evidence and analyses related to the radial mass distribution imply that the Pleiades is currently undergoing a complicated interplay of various internal dynamical effects, of which the modulation between mass segregation and binary disruption is particularly pronounced.

\end{abstract}



\section{Introduction} \label{sec:intro}

Star clusters serve as ideal laboratories for investigating the internal dynamical evolution of N-body systems. These gravitationally bound systems, comprising objects with significantly different masses and containing substantial populations of binary and multiple stars, experience various dynamical effects (see a review from \cite{2010ARA&A..48..431P}.)

Following the earliest stages of star formation and supernova activity, classical dynamical relaxation emerges as the most fundamental evolutionary driver in clusters. This process arises from interactions between objects with different masses, triggering mass segregation which consequently accelerates the escape of lower-mass objects and induces core collapse. The presence of binary/multiple stars makes the situation extremely complex. The complexity comes from the coupling of intra-binary dynamics and the overall cluster dynamics with the underlying mechanism being the energy transfer between the binary's binding energy and the kinetic energy of the third object during their interaction. Meanwhile, as individual objects, binaries will likewise obey the classical relaxation of the cluster dynamics, making them tend to converge towards the cluster center exhibiting binary segregation. 

Dynamical interactions occur frequently in high-density regions, allowing a large number of soft binaries to be disrupted quickly in the early stage \citep{2024ApJ...977..203C}. Meanwhile, new dynamical binaries also can be created through three-body interactions \citep{2010MNRAS.405..666C}. Hard binaries, on the contrary, tend to become harder during interactions over long periods of subsequent dynamical evolution \citep{1975MNRAS.173..729H}. Both the formation of new binaries and the hardening of binaries produce a heating mechanism for the cluster \citep{2011MNRAS.410.2787C}. This heating process prevents core collapse while accelerating the cluster's expansion and dissolution \citep{2014MNRAS.444...80O}. However, on the contrary, the dynamical relaxation slows the expansion and can still produce mass segregation \citep{1997MNRAS.288..749M}.

All these dynamical effects, as well as their interplay or competition, will eventually be exhibited in the radial distributions of cluster members and their mass dependence, especially for the binary population which can be characterized by the $f_{\rm b}-R$ relation. 

The evolution of the $f_{\rm b}-R$ relation is well-predicted by N-body simulations \citep{2013ApJ...779...30G,2015ApJ...805...11G}. In the early stages, binary disruption reduces the central $f_{\rm b}$ due to frequent interactions in the dense core. Over time, dynamical mass segregation causes more massive stars and hard binaries to migrate towards the center, increasing the central $f_{\rm b}$ and potentially obscuring the earlier signatures of binary disruption. In terms of observation, the binary segregation in the later stages of cluster evolution is easy to observe \citep{2024ApJ...967...44Z}, while exploring the binary disruption in the early stages is more challenging. This difficulty arises mainly from the short timescale of early disruption, which significantly reduces the probability of catching the disruption signature during this phase. Only a few studies have reported a decline in $f_{\rm b}$ towards the cluster center supporting the binary disruption, for example, in the massive young cluster NGC1818 in the Large Magellanic Cloud (LMC) \citep{2002MNRAS.331..245D,2013MNRAS.436.1497L} and in the young open cluster NGC3532 \citep{2020ApJ...901...49L} in the Milky Way. Solid observational evidence for early dynamical disruption of binaries remains scarce.

The Pleiades is a well-studied populous nearby open cluster, containing over 1400 members \citep{2019AA...628A..66L,2023AA...673A.114H}) at distances ranging from 130 to 140 pc \citep{2005A&A...429..645S,2014Sci...345.1029M}. It has already been stripped of all internal gas, leaving a typical N-body system. Although it exhibits evidence of both mass and binary segregation \citep{1998A&A...329..101R,2001AJ....121.2053A,2008ApJ...678..431C,2018A&A...612A..70O,2019PASP..131d4101G,2021AJ....162..264J,2022MNRAS.517.3525B}, it is relatively young (approximately 100 Myr \citep{2018ApJ...863...67G,2020ApJ...903...93N}) and not yet fully dynamically relaxed. As a result, it may still retain some signatures of early binary disruption.
This makes the Pleiades an ideal target for investigating the interplay of various internal dynamical effects through the mass-dependent radial distribution of cluster members, as well as the radial distribution of binaries. Possibly, one might be able to figure out how the $f_{\rm b}-R$ relation varies with time by examining binaries of different masses in this cluster \citep{2015ApJ...805...11G}. Unfortunately, despite the existence of numerous studies on binaries in the Pleiades \citep{1998A&A...329..101R,2008ApJ...678..431C,2021AJ....162..264J}, the impact of mass segregation and binary disruption processes on the $f_{\rm b}-R$ relation has not been thoroughly explored.

In this letter, by using a newly determined main sequence (MS) stellar sample, including MS single stars and unresolved MS+MS binaries in the Pleiades \citep{2025AJ....169..116L} (hereafter Paper 1), we aim to compare the radial distribution of single stars and binaries, as well as their mass dependence, to explore signatures of mass segregation and binary disruption. We then attempt to understand the underlying dynamical origins of various observational evidence. In Section~\ref{sec:sample}, we introduce the main-sequence member sample, describe the definition of stellar mass for single stars and binaries, revisit the mass functions, and evaluate the dynamical status of the Pleiades. Section~\ref{sec:da} presents and compares the mass segregation of single and binary stars, and analyzes $f_{\rm b}-R$ relations for different mass ranges. Finally, Section~\ref{sec:sum} summarizes our findings and conclusions.

\section{Main-sequence Members and Dynamical Status of the Pleiades}\label{sec:sample}

\subsection{Sample of Main-sequence Objects }\label{subsec:sample}

Our member sample for the Pleiades open cluster contains 1154 MS single stars and unresolved MS+MS binaries brighter than $G=19$ mag within a radius of 8.9$^\circ$ (Paper 1). It roughly includes $\sim 90\%$ of MS objects ranging from 0.11 to 4.9$\mathcal{M}_{\odot}$, with a binary fraction of $f_{\rm b}=0.34\pm0.02$. This sample was obtained through the following steps.

We first employed a two-component mixture model to determine the kinematic members of the Pleiades by using five-dimensional astrometric data from Gaia DR3 \citep{2016A&A...595A...1G,2023A&A...674A...1G,2023A&A...674A..32B}. Then, we used a multiband photometric technique, which is equivalent to the spectral energy distribution (SED) fitting, to determine MS objects in the cluster, including MS single stars and unresolved MS+MS binaries. Within this method, the Gaia optical and 2MASS near-infrared \citep{2006AJ....131.1163S} photometric data were adopted, and a made-to-measure empirical magnitude model for the Pleiades was built by modifying the PARSEC 1.2s theoretical model \citep{2012MNRAS.427..127B} based on the best-fitted cluster parameters from the MiMO method \citep{2022ApJ...930...44L}. We assume a priori that each cluster member contains a companion, whether or not the companion has detectable starlight, and then employed the multiband magnitudes to fit two parameters: the primary mass ($\mathcal{M}_1$) and mass ratio ($q$). A binary star is defined as having a secondary companion with its mass ($q\mathcal{M}_1$) greater than the lower mass limit in the PARSEC model (0.09$\mathcal{M}_{\odot}$). Conversely, if the mass of the secondary companion is less than this lower limit, it is considered to have a non-luminous companion and is practically defined as a single star (see Eq. 1 of Paper 1).

We have excluded $\sim 11\%$ of objects with close neighbors within 5 arcsec, which is the effective resolution of the 2MASS photometry\footnote{https://irsa.ipac.caltech.edu/data/2MASS/docs/releases/\\allsky/doc/sec2$\_$2a.html}. After the multiband fitting, $\sim 7\%$ of the remaining objects were further rejected because they did not satisfy the multiband magnitudes for a single or double MS star. These rejected objects might be MS+WD(white dwarf) binaries, multiple-star systems, and probably a few background contaminants. In short, our sample focuses exclusively on MS objects in the Pleiades. Although it is not fully complete, it extends down to the lowest mass ratios of binaries that current photometric methods can reach. 

Additionally, it is reported that the overall angular resolution of Gaia DR3 is higher than 1.5 arcsec \citep{2021A&A...649A...5F}, corresponding to $\leq$200 au at the distance of the Pleiades ($D=135$ pc or $DM_{\rm C}=5.65$ mag in Paper 1). That means, wide binaries, typically with separations $\geq$ 300 au, were all resolved as single targets, remaining only close binaries in our sample.

Our multiband fitting process is based on a Bayesian framework, so we can obtain not only the best fitting values of $\mathcal{M}_1$ and $q$ but also the posterior probability density function (PDF) in the phase space. We can therefore calculate the binary probability of each object ($P_{\rm b}$) by integrating the PDF that belongs to a binary (Eq. 5 and Table 4 of Paper 1). The distinction between singles and binaries is clear, with the effectiveness index \citep{1996AcASn..37..377S} equal to 0.88, and 1016 objects in the sample have $P_{\rm{b}} < 0.1$ or $P_{\rm{b}} > 0.9$. Nevertheless, in the following sections, we will derive statistical properties for single stars or binaries by weighting every sample object with $1-P_{\rm b}$ or $P_{\rm b}$, respectively.

\begin{figure}
	\includegraphics[width=\columnwidth]{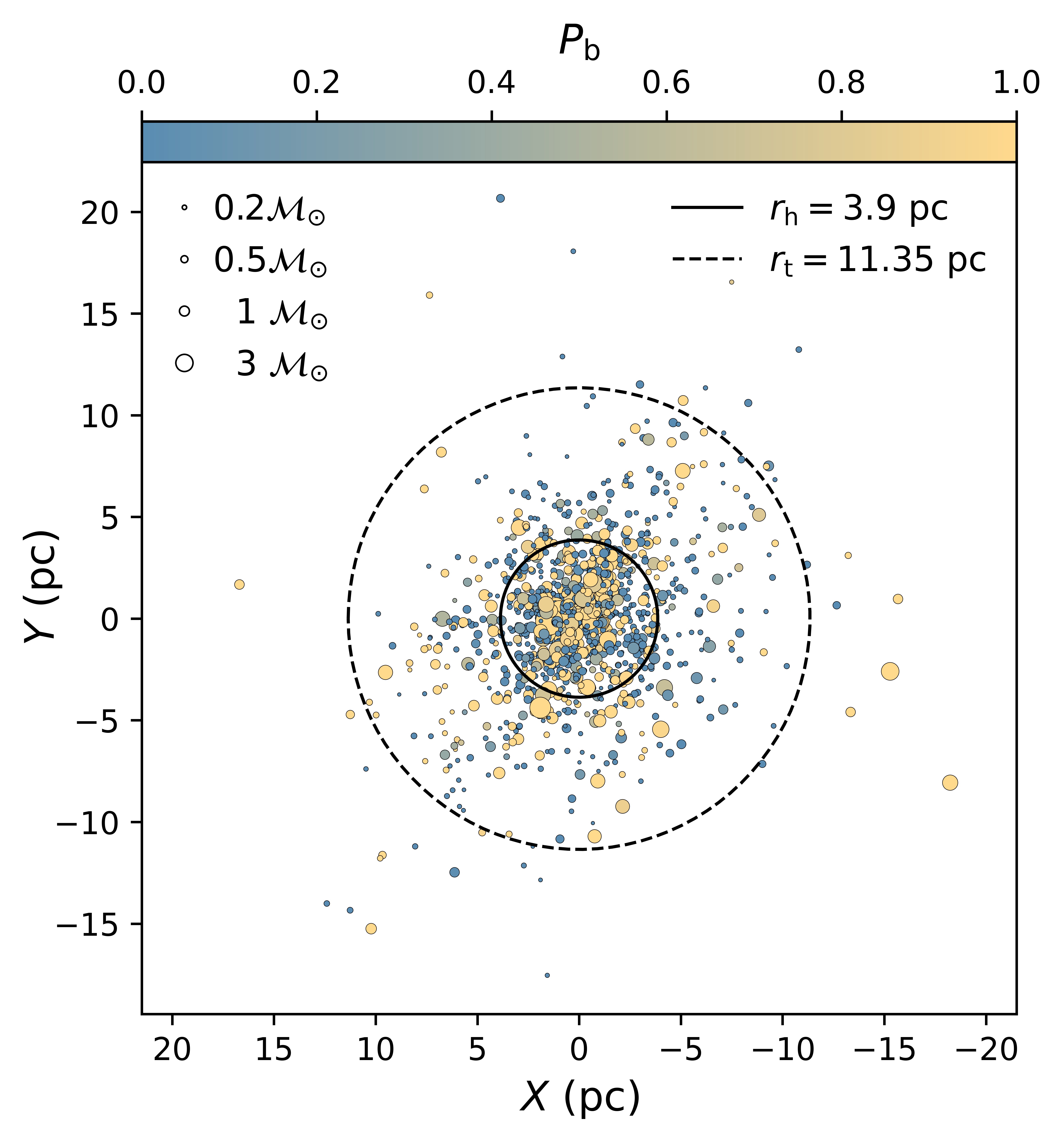}
    \caption{Distribution of main-sequence (MS) members in the Pleiades. Color represents the binary probability, and the symbol size denotes the stellar mass. The solid and dashed circles represent the half-mass radius ($r_{\rm h}$) from our sample and the tidal radius ($r_{\rm t}$) from \cite{2023A&A...677A.163A}, respectively. }
    \label{fig:data}
\end{figure}

\begin{figure}
	\includegraphics[width=\columnwidth]{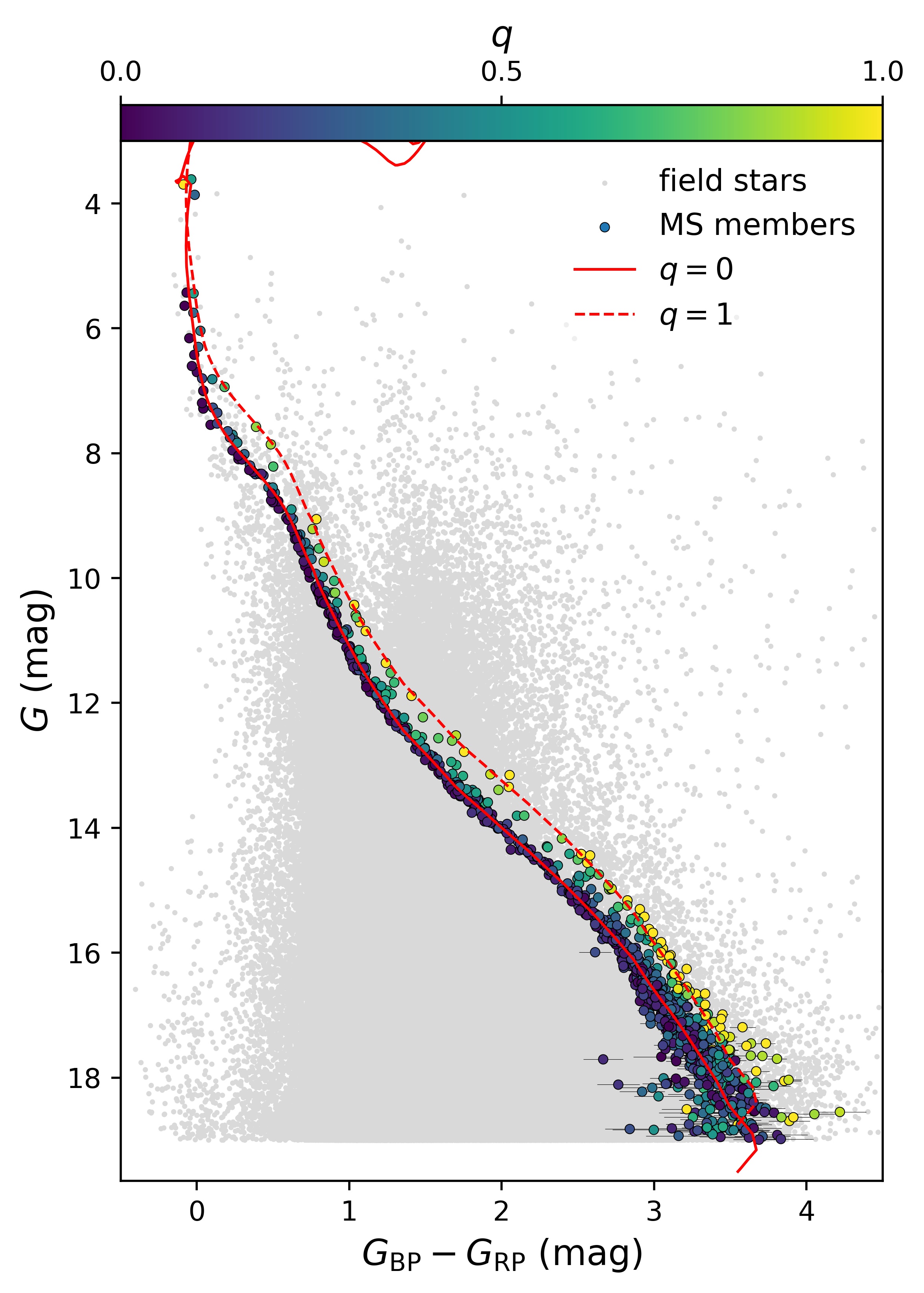}
    \caption{Color-magnitude diagram of stars in the Pleiades region for the Gaia magnitudes. The gray dots represent field stars. The circles are MS members of Pleiades, with color indicating the mass ratio. The red solid and dashed lines represent the empirical isochrones (derived from Table 3 of Paper1) for single and binary stars, respectively.}
    \label{fig:data_2}
\end{figure}

Figure~\ref{fig:data} shows that our MS sample covers the entire area within the tidal radius of the Pleiades \citep{2023A&A...677A.163A}. It is also clear that the central part is more concentrated and almost spherically symmetric. While the outskirt is elongated, indicating the consequences of the Galactic tidal force \citep{2019A&A...628A..66L,2021A&A...645A..84M,2021RNAAS...5..173L,2022MNRAS.517.3525B}. Figure~\ref{fig:data_2} displays sample stars in the CMD of Gaia magnitudes, colored by their mass ratios. One can find that the final sample of MS population matches the empirical magnitude model very well.

\subsection{Stellar Mass and Mass Function}\label{sec:mf}

Our dynamical analysis in this work is based on the total stellar mass of individual member objects, or the nominal binary system, rather than the primary mass ($\mathcal{M}_1$) of a binary alone. The total mass of a binary $\mathcal{M}_{\rm obj}=\mathcal{M}_1+q\mathcal{M}_1$, where $\mathcal{M}_1$ and $q$ are taken with their best fitting values listed in Table 4 of Paper 1. For the mass of a single star, we used the same calculation method as for binaries. That means, we not only considered the MS star itself but also assumed that it contains a non-luminous companion, a brown dwarf or a giant planet for instance, with a mass of $q\mathcal{M}_1$. According to Section 3.2  and Eq. 1 of Paper 1, this secondary mass should be randomly assigned from zero to 0.09$\mathcal{M}_{\odot}$. While this is not true for individual single stars, it is a good approximation of what is more reasonable from a statistical point of view.

\begin{figure}
	\includegraphics[width=\columnwidth]{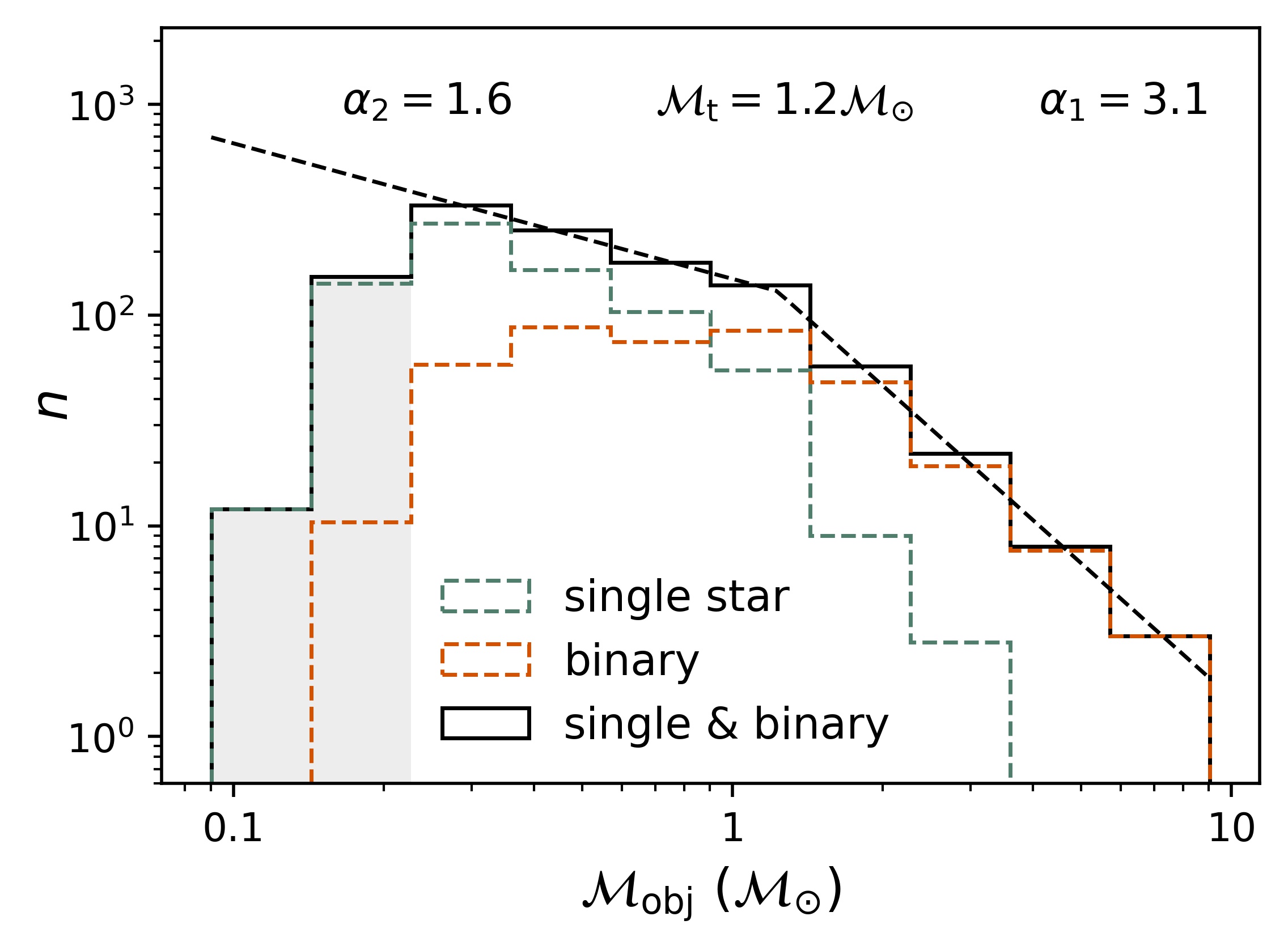}
    \caption{Mass function of MS members. The black, green, and orange histograms show the mass distribution of all objects, single stars, and binaries, respectively. The objects in the shaded region are incomplete. The black dashed line shows the best-fit mass function for the complete part with the format of $\mathcal{M}^{1-\alpha}$. }
    \label{fig:mf}
\end{figure}

In our sample, the lowest mass of a single star is 0.11$\mathcal{M}_{\odot}$, so the sample of $\mathcal{M}_{\rm obj} < 0.22\mathcal{M}_\odot$ is not complete. Figure~\ref{fig:mf} shows the $\mathcal{M}_{\rm obj}$ distributions of our sample. For the mass of all objects (black histogram), we find that the mass function for the complete part ($\mathcal{M}_{\rm obj} > 0.22\mathcal{M}_\odot$) follows a two-segment power-law profile, which is similar to the first two brighter domains of the primary mass function obtained by  \cite{2015A&A...577A.148B}. Both segments can be uniformly expressed as $\mathcal{M}^{1-\alpha} $, and the best fitting values are $\alpha_1=3.1, \alpha_2=1.6$ for the high-mass and low-mass ranges respectively, with a transition mass of $\mathcal{M}_{\rm t}=1.2 \mathcal{M}_{\odot}$. 

Notably, the mass functions of singles and binaries are significantly different. Binaries are more likely distributed in the higher mass region, leading to the gradual increase of $f_{\rm b}$ towards the massive stars, which is also found and exhibited in Figure 11 of Paper 1. This phenomenon reminds us that, when we discuss binary distributions,  the influence of their stellar mass has to be taken into account.

\subsection{Dynamical Status} \label{sec:params}

Assuming our sample can represent the radial mass distribution profile of the cluster, we estimated the projected half-mass radius of the Pleiades, $R_{\rm h}=2.9~\text{pc}$. Accounting for the projection effect, we multiplied it by a factor of 4/3 \citep{1987degc.book.....S} and obtained the deprojected half-mass radius $r_{\rm h}=3.9~\text{pc}$, which is shown by the solid line in Figure~\ref{fig:data}. 
 
The total object number of our sample is 1154, and the sum of their $\mathcal{M}_{\rm obj}$ is 737$\mathcal{M}_\odot$.  Given the $\sim 10\%$ incompleteness of our sample, and the fact that the sample does not include small objects less than 0.11$\mathcal{M}_\odot$, as well as those multiple star systems and MS+WD binaries that have been eliminated, the total object number and mass of the Pleiades might be $N_{\rm C}\sim1450$ and $\mathcal{M}_{\rm C} \sim 830 \mathcal{M}_\odot$. These values are in line with various results of previous works \citep{2019PASP..131d4101G,2020AstBu..75..407D,2022MNRAS.516.5637E,2023MNRAS.525.2315A}. For instance, \cite{2022MNRAS.516.5637E} reported that $N_{\rm C}\sim1448$ and $\mathcal{M}_{\rm C} \sim 807 \mathcal{M}_\odot$, which is in good agreement with our estimation. The average object mass is then estimated as $\overline{\mathcal{M}} \sim  0.57\mathcal{M}_\odot$.

The ratio of the cluster's age to its relaxation time indicates its dynamical age. Following \cite{1969ApJ...158L.139S}, the half-mass relaxation time of the cluster is:
\begin{equation}
t_{\rm rh} =\frac{0.17N_{\rm C}}{\ln(\lambda N_{\rm C})}\sqrt{\frac{r_{\rm h}^3}{G\mathcal{M}_{\rm C}}}, 
\end{equation}
where $\lambda$ is a constant for which we use a value of 0.1 \citep{1994MNRAS.268..257G}. Then the $t_{\rm rh}$ of the Pleiades is calculated as 195 Myr. Given the cluster age of 106 Myr (Paper1), we obtained the ratio to be $\sim 0.54$, suggesting that the dynamical relaxation of the Pleiades is rather inadequate, although it has already passed about 7 crossing times, with $t_{\rm cr}\approx 15$ Myr given by \cite{1998MNRAS.299..955P}.

For the system that has not fully relaxed, one can also estimate a mass-segregation timescale for stars
of a given mass, following \cite{1969ApJ...158L.139S}: 
\begin{equation}
t_{\rm seg} = \frac{\overline{\mathcal{M}}}{\mathcal{M}_{\rm seg}} t_{\rm rh},
\end{equation}
where $\mathcal{M_{\rm seg}}$ is the mass of the object of interest. Taking the above values of $t_{\rm rh}$ and $\overline{\mathcal{M}}$, and the age of the Pleiades, we can find the current $\mathcal{M_{\rm seg}} \sim 1 \mathcal{M_\odot}$. Although this is a very rough estimation, it offers a fiducial reference. That is, those cluster members more massive than this level will be expected to have significant dynamical mass segregation.

\section{Radial Distributions} \label{sec:da}

\subsection{Mass Segregation}\label{sec:m_r}

\begin{figure}
    \includegraphics[width=\columnwidth]{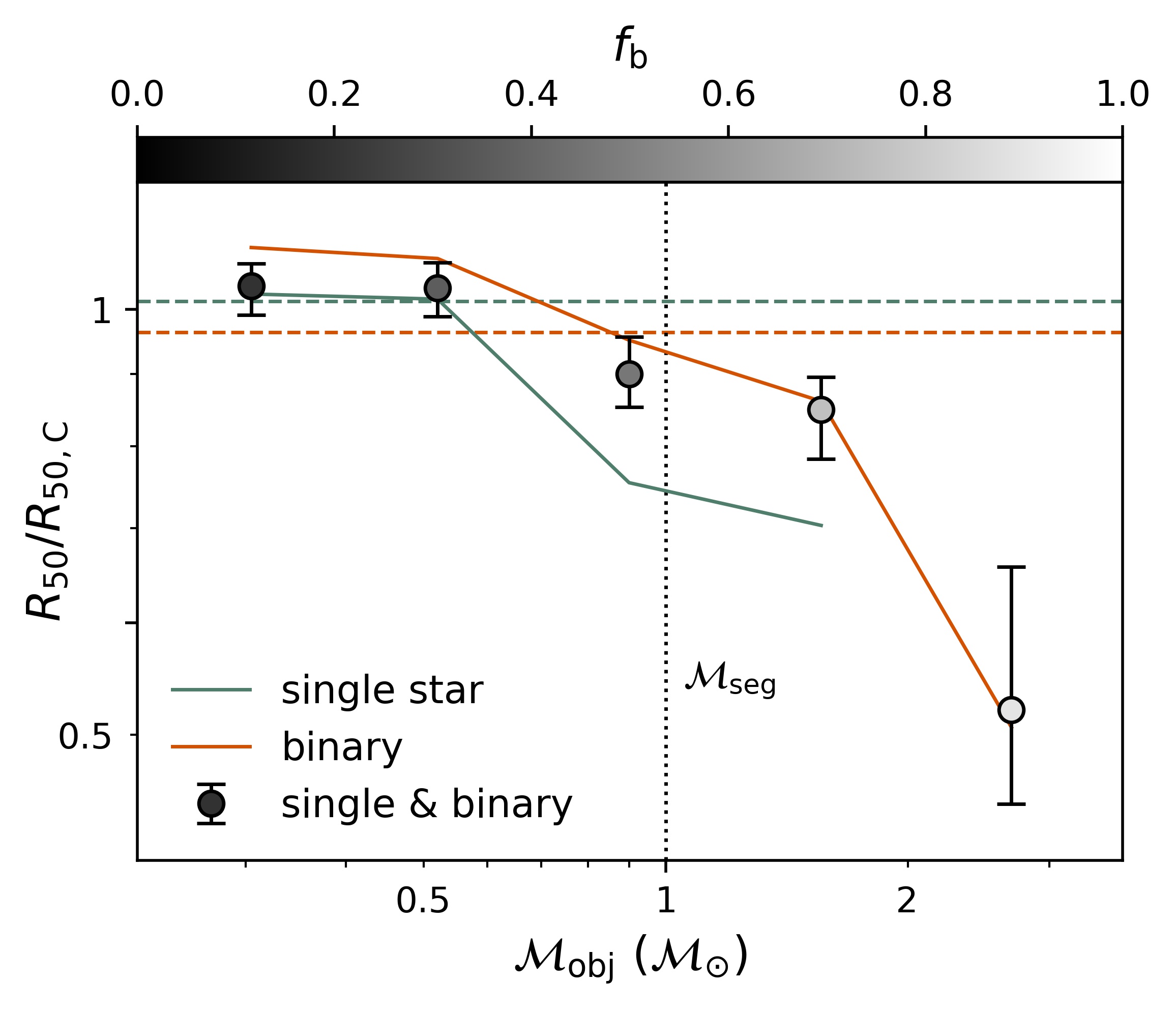}
    \caption{$R_{50}-\mathcal{M}_{\rm obj}$ relations for all objects (circles with error bars), single stars (green line), and binaries (orange line). Mass bins are separated by [0.38, 0.66, 1.14, 2.00]$\mathcal{M}_{\odot}$, which are approximately equal logarithmic-mass intervals except for the most massive bin. The most massive bin for single stars is neglected due to insufficient numbers. $R_{\rm 50,C}$ is the half-number radius of the whole MS sample. The error bars are estimated from the interval of 16\% to 84\% of the corresponding $R$ distribution of each subsample, but divided by the factor of $n_{\rm sub}^{1/2}$. The gray degree of symbols represents the binary fraction in each mass bin. Horizontal green and orange dashed lines represent $R_{\rm 50}$ values of singles and binaries for the entire mass range. The vertical black dotted line denotes the location of $\mathcal{M}_{\rm seg}$.}
    \label{fig:m_r}
\end{figure}

We divided the sample of $\mathcal{M}_{\rm obj}>0.22\mathcal{M}_{\odot}$ into five mass bins and compared the half-number radii ($R_{50}$) of subsamples as functions of object mass to identify mass segregation. A smaller $R_{50}$ indicates a higher concentration towards the center. As shown in Figure~\ref{fig:m_r},  the $R_{50}$ of five subsamples for all objects (circles with error bars), which contain both singles and binaries, gradually decrease with mass, indicating significant mass segregation in this cluster.  The mass segregation of the Pleiades has already been found and confirmed in many previous works, using different data and methods. \citep{1998A&A...329..101R,2001AJ....121.2053A,2008ApJ...678..431C,2018A&A...612A..70O,2019PASP..131d4101G,2021AJ....162..264J,2022MNRAS.517.3525B}. It has also been validated by dedicated N-body simulations for the Pleiades \citep{2004A&A...426...75M}, while \cite{2010MNRAS.405..666C} even claimed that the Pleiades already had significant mass segregation just after the molecular cloud gas was expelled.   

In this work, we should further note two other observational features. One is that the mass segregation is more evident for objects more massive than the $\mathcal{M}_{\rm seg}$, but less obvious when object mass is below this criterion. This phenomenon is consistent with the concept of $\mathcal{M}_{\rm seg}$ in the classical dynamical scenario, although we cannot fully discount the influence of binary heating or completely rule out the primordial mass segregation. Another feature is that, if we consider the singles and binaries separately (green and orange lines), we still observe strong evidence of mass segregation for both of them. This feature indicates that the present mass segregation concerns only the object mass and is not relevant to whether they are single or binary stars.

Interestingly, a comparison of the $R_{\rm 50}$ of singles and binaries within the same mass range reveals that the singles are more concentrated than binaries across all mass ranges except the most massive bin, which consists almost entirely of binaries. This is the first time such a comparison has been addressed. Despite the non-negligible uncertainties in $R_{50}$, the overall difference between singles and binaries remains definite. To explain this phenomenon, we note that dynamical mass segregation alone offers no reason for single stars to preferentially converge towards the center over binaries. Therefore, the remaining explanation is that it is a consequence of the fact that binaries are more easily disrupted at the cluster center. So we interpret this phenomenon as an obvious signature of the dynamical disruption of binaries in the Pleiades.  This process appears to be ongoing, as frequent interactions among cluster members would otherwise quickly erase this difference. Another implication of this result is that, while new binaries may be created in the central region, this does not fully compensate for the reduction due to disruption, at least for the lower-mass close binaries in the current sample.

In contrast, if we consider the entire sample, the $R_{\rm 50}$ of binaries (orange dashed line) is smaller than that of singles (green dashed line), which was also reported in previous works \citep{1998A&A...329..101R,2021AJ....162..264J}. This has been canonically explained by the higher average mass of the binary systems or the primordial concentration of binaries in young clusters, just as claimed by \cite{2024ApJ...962L...9M} for another young OC M35. However, by combining the facts that the mass function of binaries is more top-heavy than that of single stars, and the increase of $f_{\rm b}$ along with object mass, we can understand that this phenomenon might be simply due to the dynamical mass segregation. That is, along with the sinking of massive objects that contain a higher fraction of binaries, more binaries naturally converge towards the cluster center.

\subsection{Radial Distribution of the Binary Fraction}\label{sec:fb_r}

\begin{figure}
	\includegraphics[width=\columnwidth]{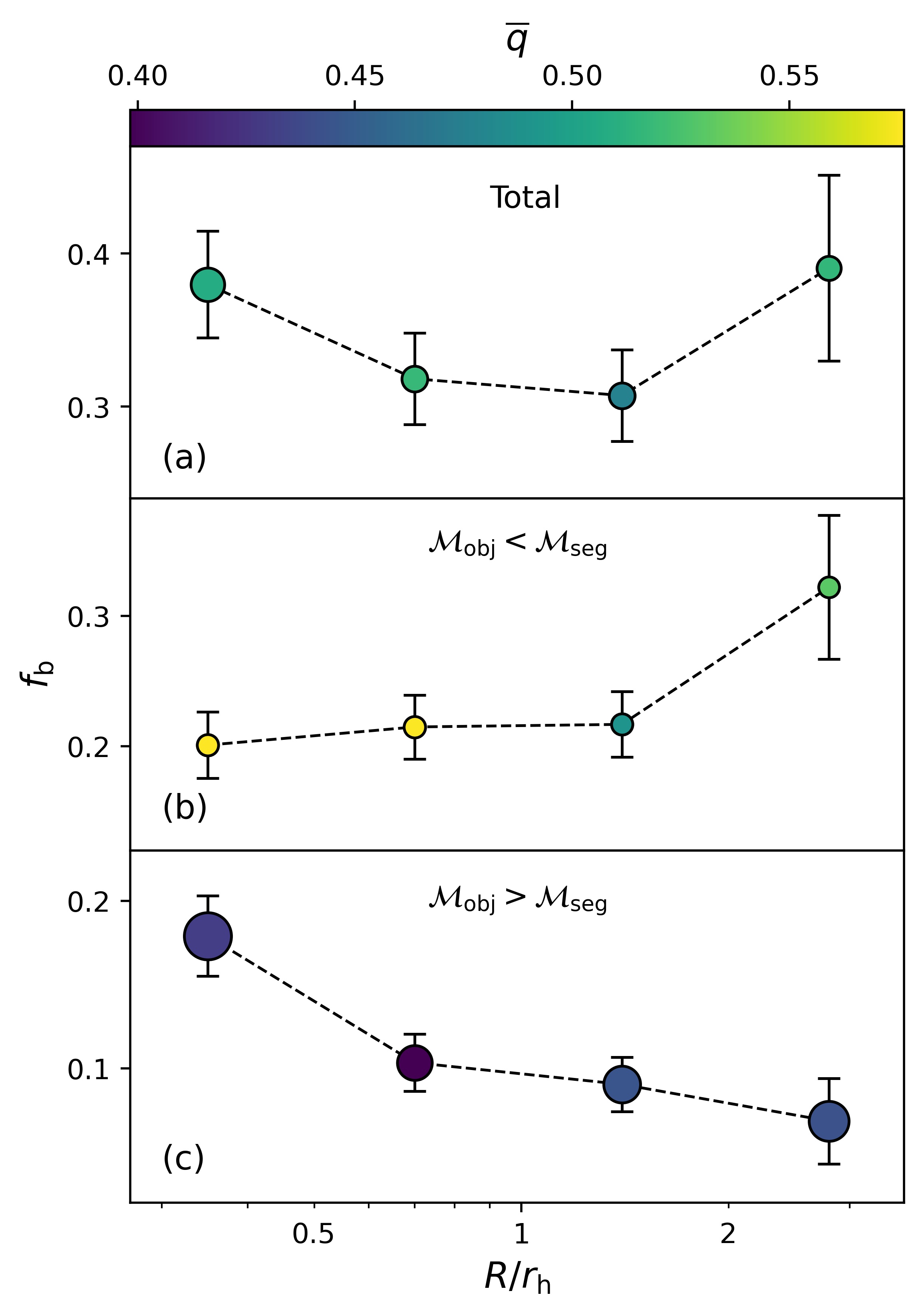}
    \caption{$f_{\rm b}-R$ relation in different mass ranges within the tidal radius. Panel (a) shows the relation for all binaries. Panels (b) and (c) present the relations for low and high mass binaries separated by $\mathcal{M}_{\rm seg}$. The size and color of symbols represent the average mass and average mass ratio in each $R$ bin, respectively. The errors are estimated based on the Poisson fluctuation.} 
    \label{fig:fb_r}
\end{figure}

The radial distribution of $f_{\rm b}$ reflects the combined effect of binary disruption, mass segregation, and dynamical formation of new binaries. We divided the distance to the center into four bins, separated by [0.5, 1.0, 2.0] $r_{\rm h}$, and calculated the $f_{\rm b}$ for each bin. Figure~\ref{fig:fb_r} shows the $f_{\rm b}$ as function of radius. Panel (a) shows the result of the whole sample. It exhibits a bimodal distribution with higher binary fractions near the center and in the outermost region, while the lower binary fractions appear around the $r_{\rm h}$. It is also found that the typical mass of inner binaries is larger than that of the outer binaries (see the symbol size in panel (a)), implying that the $f_{\rm b}-R$ relation is mass-dependent.

The origin of such a non-monotonic distribution can be understood when we divide the sample by stellar mass, and consider their $f_{\rm b}-R$ relation separately. In the range of $\mathcal{M}_{\rm obj} < \mathcal{M}_{\rm seg}$, where stars have no obvious mass segregation, one may find that $f_{\rm b}$ increases with radius. This trend is consistent with the consequences of the dynamical binary disruption since the inner part, having more frequent interactions, will raise the probability of disruption. Another indirect evidence supporting this mechanism is that, on average, the mass ratios ($q$) of two inner bins are larger than those of outer ones. That is because the binary binding energy $E_{\rm b} \propto q\mathcal{M}^2_1$. So lower-$q$ binaries with lower binding energy are more easily disrupted. 

On the other hand, for the subsample of $\mathcal{M}_{\rm obj} > \mathcal{M}_{\rm seg}$, $f_{\rm b}$ gradually decreases with radius. This phenomenon could be explained by the significant mass segregation effects of these massive objects. In the segregation process, more massive objects, shown as the largest symbol in Panel (c),  sink to the cluster center earlier. Since the more massive objects bring higher fractions of binaries, when they gather to the center of the cluster at an earlier time, they will also significantly increase the $f_{\rm b}$ value. 

If we assume that our current sample does include some newly created close binaries, it is not difficult to understand that these binaries form mainly in the center of the cluster, and in fact, massive stars are more likely to capture a companion. This process would further increase the fraction of massive binaries in the central region. Thus, it could also be one of the reasons shaping the radial distribution of massive binaries.

The $f_{\rm b}-R$ relation of the Pleiades has not been discussed in previous works. However, a similar distribution has been found in NGC 1805, which is a more massive young star cluster in the LMC \citep{2013MNRAS.436.1497L}. This trend of $f_{\rm b}$ decreasing with increasing radius in the central region, followed by a slow increase, can also be predicted by the N-body simulation. \cite{2013ApJ...779...30G} reported that in the simulation of massive clusters, after roughly one initial $t_{\rm rh}$, the radial binary frequency distribution becomes bimodal since the innermost binaries have already segregated toward the core. 

In brief, the $f_{\rm b} - R$ relation of the Pleiades reveals both signatures of binary disruption and mass segregation, which lead to higher values of $f_{\rm b}$ in the outer and inner parts, respectively.

\section{Conclusions}\label{sec:sum}

By using a well-defined MS sample of the Pleiades, we revisited the object mass of cluster members and found that the binary's mass function is more top-heavy than that of the single stars. 

After diagnosing the various observational phenomena associated with the radial mass distribution of the Pleiades, we concluded that all observed distributions of single stars and binaries can be attributed to the complicated internal dynamical processes within the cluster, and mainly reflected in the effects of binary disruption and mass segregation, and their interplay.

Mass segregation is directly evident in the concentration of massive cluster members ($\mathcal{M}_{\rm obj} > \mathcal{M}_{\rm seg}$), observed for both single stars and binaries. Another indirect indicator of mass segregation exists in the gradual increase of binary fractions toward the cluster center within the higher-mass subsample.

Two signatures of binary disruption are also apparent. First, within given mass ranges, binaries tend to be distributed farther from the center than single stars. Second, the lower-mass subsample exhibits a monotonic increase in binary fraction towards the outskirts.

Furthermore, we argue that there is no compelling need to invoke a primordial concentration of binaries in the Pleiades. While we cannot completely rule out the initial segregation, an early dynamical mass segregation process could have efficiently transported massive members with high binary ratios, including those newly formed dynamical binaries, to the center of the cluster.

This work highlights the importance of the radial distribution of binary fractions. It plays a critical role in diagnosing various dynamical contributions, not only the classical dynamics but also the dynamics of binaries in a cluster.

\begin{acknowledgments}
We thank the referee for constructive suggestions on the motivation of this paper. This work was supported by the National Natural Science Foundation of China (NSFC) under grants 12273091 and 12303026; the Science and Technology Commission of Shanghai Municipality (Grant No.~22dz1202400); the science research grants from the China Manned Space Project with No.~CMS-CSST-2021-A08; This work was also sponsored by the Young Data Scientist Project of the National Astronomical Data Center and the Program of Shanghai Academic/Technology Research Leader. This work has made use of data from the European Space Agency (ESA) mission Gaia (https://www.cosmos.esa.int/gaia), processed by the Gaia Data Processing and Analysis Consortium (DPAC; https://www.cosmos.esa.int/web/gaia/dpac/consortium). Funding for the DPAC has been provided by national institutions, in particular, the institutions participating in the Gaia Multilateral Agreement. This work made use of data products from the Two Micron All Sky Survey, which is a joint project of the University of Massachusetts and the Infrared Processing and Analysis Center/California Institute of Technology, funded by the National Aeronautics and Space Administration and the National Science Foundation. 

\end{acknowledgments}




\bibliography{reference}{}
\bibliographystyle{aasjournal}



\end{document}